\documentclass[a4paper]
{article}

\usepackage[utf8]{inputenc}
\usepackage[T1]{fontenc}

\usepackage[a4paper,margin=2.5cm,right=3cm]{geometry}
\usepackage{pdflscape}
\usepackage{lmodern}

\usepackage{mathtools,amsmath, amssymb,amsthm}
\usepackage{tensor,mathrsfs,upgreek,stmaryrd}
\usepackage[normalem]{ulem}
\usepackage[makeroom]{cancel}

\usepackage{cite, aas_macros}
\usepackage[sort&compress,numbers]{natbib}

\usepackage[dvipsnames]{xcolor}
\usepackage{url}
\usepackage[colorlinks=true, citecolor=Blue, linkcolor=Red]{hyperref}

\let\t\tensor
\let\p\partial

\newcounter{mnotecount}[section]

\newtheorem{proposition}{Proposition}[section]

\newcommand{\g}{\gamma}
\newcommand{\m}{\mathfrak{m}}
\newcommand{\redmusquared}{\mu^2_{\mathrm{P}}}

\let\t\tensor
\let\p\partial

\newcommand{\hyp}{\mathscr{S}}

\def\R{\mathbb R}

\def\mcL{{\mathcal L}}
\def\m{\mathfrak m}

\newcommand{\A}{{\tilde A}}
\newcommand{\tnabla}{{\tilde \nabla}}
\newcommand{\tBox}{{\tilde \Box}}
\newcommand{\tR}{{\tilde R}}

\newcommand{\redp}{{\lambda}}

\usepackage{textgreek}
\def\permeability{\text{\textmu}}
\def\permittivity{\text{\textepsilon}}

\usepackage{authblk}

\numberwithin{equation}{section}

\begin{document}
\title{The Cauchy problem for the Proca equation in gravitating dielectric media%
\thanks{Preprint UWThPh-2023-17}}

\author[1,2]{F.~Steininger}
\author[2]{P.~T.~Chruściel}
\affil[1]{\footnotesize University of Vienna, Faculty of Physics, Vienna Doctoral School in Physics, Boltzmanngasse 5, 1090 Vienna, Austria}
\affil[2]{\footnotesize University of Vienna, Faculty of Physics and Research platform TURIS, Boltzmanngasse 5, 1090 Vienna, Austria}

\date{\today}
\maketitle

\begin{abstract}
We analyze the Cauchy problem for the Proca equation in gravitating dielectric media.
\end{abstract}

\section{Introduction}

There is continuing interest in the physics community about a potential mass for photons, with generation mechanisms proposed in~\cite{BONETTI2017203,AdelbergerDvaliGruzinov}.
In a recent note~\cite{ChMS} we have shown that the Proca equation is incompatible with experimental evidence for propagation of Maxwell fields in coaxial cables,
 \emph{no matter how small the Proca mass is, if non zero}. This result invokes boundary conditions corresponding to a perfectly conducting medium, which might be viewed as a an overly idealised situation. One is then led to the question of  the properties of the Proca equation in situations where no such boundary conditions arise.

In view of   recent work on gravitational effects in waveguides~\cite{HMMMCW,BCHKW}, there is a need to consider generalisations of the Proca equation in gravitating dielectric media. The object of this note is to analyse the Cauchy problem for the Proca equation in three situations: a) Proca fields in vacuum in a gravitational field (see Proposition~\ref{P8V22.1} below with $n=1$); b) a seemingly natural generalisation of the Proca equation for a gravitating dielectric medium  (cf.\ Proposition~\ref{P8V22.1} below with $n\ne 1$); c) another  seemingly natural generalisation of the Proca equation for dielectrics in Minkowski spacetime (addressed in Proposition~\ref{p25IV23.1} below with $n\ne 1$).

\section{The model}

Now, in the absence of currents,
the curved-spacetime model for massive photons is obtained by adding a mass term to the Maxwell Lagrangian, as
\begin{equation}
	\mcL =
 \Big(
 - \frac{1}{4} \t{ g }{^\alpha^\beta} \t{ g }{^\rho^\sigma} \t{ F }{_\alpha_\rho} \t{ F }{_\beta_\sigma} -
  \frac 12 \frac{\redmusquared c^2}{\hbar^2}  \t{ g }{^\alpha^\beta} A_\alpha A_\beta 
  \Big)
   \sqrt{|\det g|}
  \,,
\label{eq:MaxwellL}
\end{equation}
with
\begin{equation}
	F_{\alpha \beta} = \p_\alpha A_\beta - \p_\beta A_\alpha
	\,.
\end{equation}
Here $A_\sigma$ is the photon potential, $g_{\alpha \beta}$ a general spacetime metric and  $\nabla_\alpha$ is the covariant derivative with respect to $g_{\alpha \beta}$. Henceforth we will use units so that $\hbar = c = 1$.

As already hinted-to, there exist two obvious natural  generalisations of Proca equation to gravitating dielectric media, arising from the Lagrangian
\begin{equation}
	\mcL =
 \Big(
 - \frac{1}{4 \permeability} \t{ \gamma }{^\alpha^\beta} \t{ \gamma }{^\rho^\sigma} \t{ F }{_\alpha_\rho} \t{ F }{_\beta_\sigma} -
  \frac 12 \redmusquared  \t{ \m }{^\alpha^\beta} A_\alpha A_\beta
  \Big)
   \sqrt{|\det g|}
  \,,
\label{eq:L}
\end{equation}
with   $\m^{\alpha\beta} \in \left\{ g^{\alpha \beta}, \g^{\alpha \beta}   \right\}$ in the mass term. Here $\t\gamma{^\alpha^\beta}$ is Gordon's optical metric~\cite{Gordon23},
\begin{equation}
	\t\gamma{^\alpha^\beta}
		= \t{g}{^\alpha^\beta}
		+ (1 - n^2) \t u{^\alpha} \t u{^\beta}\,,
		\label{eq:Gordon}
\end{equation}
where $u^\beta$ is the four-velocity of the dielectric medium  normalized to $g_{\mu\nu}u^\mu u^\nu = -1$, and $n \equiv \sqrt{ \permittivity \permeability} > 0$ is the refractive index with permittivity $\permittivity$ and permeability $\permeability$,
assumed to be  a smooth function on  spacetime.
 The physical interpretation of~\eqref{eq:L}-\eqref{eq:Gordon} is that the medium is linear,  locally isotropic and non-conducting.
 Additionally we take the material to be non-magnetic, i.e. $\permeability = 1$, to avoid unnecessary clutter.

Yet another possibility, which we will also   explore, and which  does not introduce physically-unmotivated fields in the equations, is to take
\begin{equation}\label{22IV23.1}
\m^{\alpha\beta} = g^{\alpha\beta} + \redp u^\alpha u^\beta
 \,,
 \quad
  \redp \in \R
  \,,
\end{equation}
so that the case  $\m^{\alpha\beta}= \g^{\alpha \beta}$ becomes a special case $\redp =1-n^2$ of \eqref{22IV23.1}, while $\m^{\alpha\beta}=  g^{\alpha \beta}$ becomes \eqref{22IV23.1} with $\redp =0$.

The Euler-Lagrange equations are

\begin{equation}
	\nabla_\alpha \t{ G }{^\alpha^\beta} - \redmusquared  \t{ \m }{^\alpha^\beta} A_\alpha = 0\,,
\label{eq:fieldeq}
\end{equation}
where $\t{G}{^\alpha^\beta} := \t{ \g }{^\alpha^\rho}\t{ \g }{^\beta^\sigma} \t{ F }{_\rho_\sigma}$. Since
\begin{equation}\label{eq:detg}
  \det g_{\alpha\beta} = n^{2}\det \g_{\alpha\beta}
  \,,
\end{equation}
 it is convenient to rewrite~\eqref{eq:fieldeq}  as
\begin{eqnarray}
	 0 & = & \nabla_\alpha \t{ G }{^\alpha^\beta} - \redmusquared  \t{ \m }{^\alpha^\beta} A_\alpha
	 \\\notag
 & = &
 \frac{1 }{\sqrt{|\det g|}}
 \p_\alpha (\sqrt{|\det g|} \t{ G }{^\alpha^\beta}) - \redmusquared  \t{ \m }{^\alpha^\beta} A_\alpha
 \nonumber
\\
  & = &
 \frac{1 }{n }
 \tnabla_\alpha (n \t{ G }{^\alpha^\beta}) - \redmusquared  \t{ \m }{^\alpha^\beta} A_\alpha
    \,,
\label{eq:fieldeqoptical}
\end{eqnarray}
where $\tnabla$ is the covariant derivative with respect to the metric $\g$.

Asuming that $\redmusquared\ne 0$, the application of the ``optical covariant derivative'' $\tnabla$ to \eqref{eq:fieldeqoptical} leads to

\begin{align}
	0 &= \tnabla_\beta \left( n \m^{\alpha \beta} A_\alpha \right)
	 = n \nabla_\beta \left( \m^{\alpha \beta} A_\alpha \right)
 \,,
	\label{eq:constraint}
\end{align}
where symmetry of the Ricci tensor has been used.

We wish to show that the equations above have a well posed Cauchy problem. For this let $\hyp$ be a spacelike hypersurface given by the equation $\hyp=\{x^0=0\}$.
A Gauss-type constraint arises from the fact that \eqref{eq:fieldeqoptical} with $\beta=0$ does not contain time-derivatives of $G^{\mu\nu}$:
\begin{equation}
  n \redmusquared \t{ \m }{^\alpha^0} A_\alpha
    =
    \tnabla _i ( n \t{ G }{^i^0})
\label{10III22.3cd}
\end{equation}
(if   $\redmusquared =0$ this is of course the Gauss constraint equation, keeping in mind that there are no exterior currents or charges in the model under consideration).

Continuing further, \eqref{eq:fieldeqoptical} can be restated as a wave equation for $A_\alpha$.
Defining
\begin{equation}
 \A_\alpha := n A_\alpha\,,
\end{equation}
and
 using
\begin{equation}
 F_{ \alpha \beta }=2 \partial_{ [\alpha } A_{\beta]}
  = 2 \tnabla_{ [\alpha } A_{\beta]}
  \,,
\end{equation}
one obtains

\begin{align}
	0
\notag	&= 2 \g^{ \alpha \sigma }  \tnabla_\alpha (\tnabla_{ [ \sigma } (n A_{ \nu] }) + A_{ [ \sigma } \tnabla_{ \nu ]} n ) - \redmusquared  \m^{ \alpha \beta } \g_{ \beta \nu } \A_\alpha
		\\
	&=
  \tBox_\g \A_\nu
   -
    \g^{\alpha \beta} \tR_{ \alpha \nu } \A_\beta - \tnabla_\nu (\g^{ \alpha \beta } \tnabla_\alpha \A_\beta)
		+ 2 \g^{ \alpha \beta }  \tnabla_\alpha ( \A_{ [ \beta } \tfrac{1}{n} \tnabla_{ \nu ]} n ) - \redmusquared \m^{ \alpha \beta } \g_{ \beta \nu } \A_\alpha\,,
	\label{5X22.2}
\end{align}
where  $ \tBox_\g = \g^{ \alpha \beta } \tnabla_\alpha \tnabla_\beta $, and $\tR _{\alpha\beta}$ is the Ricci tensor of the metric $\g_{\alpha\beta}$.

 \section{The case $\m = \g$}
Imposing the constraint \eqref{eq:constraint}, we note that \eqref{5X22.2} now reads
\begin{equation}
	0 =  \tBox_\g \A_\nu
 -
  \g^{\alpha \beta} \tR_{ \alpha \nu } \A_\beta
  		- \g^{\alpha\beta} (\tnabla_\alpha \A_\nu)\tfrac{1}{n}\tnabla_\beta n
		+ 2 \g^{ \alpha \beta }  \tnabla_\alpha ( \tfrac{1}{n} \tnabla_{[\nu } n)\A_{ \beta] }
		- \redmusquared \A_\nu\,.
\label{5X22.3}
\end{equation}
Any solution of \eqref{eq:fieldeq} must satisfy both this equation and \eqref{eq:constraint}. Moreover, every solution of this equation which also satisfies \eqref{eq:constraint} will solve \eqref{eq:fieldeq}.
So the strategy is to use \eqref{5X22.3}   as an evolution equation for $\A_\nu$, and to isolate these solutions of \eqref{5X22.3} which satisfy \eqref{eq:fieldeq}.

Now, solutions of \eqref{5X22.3}   are parameterised by their Cauchy data $(\A_\alpha, T^\nu \tnabla_\nu \A_\alpha)\big|_\hyp $,  where $T^\nu$ is e.g.\  a timelike unit normal vector field orthogonal to $\hyp$.
These Cauchy data are not arbitrary if we want to obtain a solution satisfying \eqref{eq:fieldeq}. Indeed, in Gauss coordinates near $\hyp$ for the metric $\g_{\alpha\beta}$ (so that   $\g_{0i}\equiv 0$),   Equations~\eqref{eq:constraint}  and \eqref{10III22.3cd} can be rewritten as

\begin{eqnarray}
  \g^{00}\tnabla _0 \A_0\big|_\hyp
    &=&
      - \g^{ij}\tnabla _i \A_j
   \,,
    \label{8V22.31}
\\
  \g^{ij} \tnabla _i\tnabla _0 \A_j \big|_\hyp
   &=&
    \big(
  \g^{ij}\tnabla_i \tnabla _j -   \redmusquared
  \big) \A_0
    + 2 \g^{ij} \tnabla_i (\A_{[j} \tfrac{1}{n}\tnabla_{0]} n)
   \,,
    \label{8V22.32a}
\end{eqnarray}
where the second equation can also be rewritten as
\begin{eqnarray}
\tnabla _0 (\g^{ij}\tnabla _i \A_j) \big|_\hyp
   &=&
    \big(
  \g^{ij}\tnabla_i \tnabla _j -   \redmusquared
  \big) \A_0
  -
   \g^{\mu \nu } \tR_{\mu 0} \A_\nu
    + 2 \g^{ij} \tnabla_i (\A_{[j} \tfrac{1}{n}\tnabla_{0]} n)
   \,.
    \label{8V22.32b}
\end{eqnarray}
These equations determine   $\tnabla _0 \A_0\big|_\hyp $ and  $\tnabla _0 (\g^{ij}\nabla _i \A_j)\big|_\hyp $ in terms of the remaining data.

Consider, thus, a solution of \eqref{5X22.3} with Cauchy data satisfying \eqref{8V22.31}-\eqref{8V22.32a} in adapted coordinates as above.
Clearly the function  $\tnabla _\alpha \left( \t{ \g }{^\alpha^\beta} \A_\beta \right) $ satisfies  $\tnabla _\alpha \left( \t{ \g }{^\alpha^\beta} \A_\beta \right)\Big|_\hyp =
0$ by \eqref{8V22.31}. Next, a simple calculation shows that

\begin{eqnarray}
\tnabla _0 \tnabla_\alpha \left( \t{ \g }{^\alpha^\beta} \A_\beta \right)\Big|_\hyp
 & = &
   \t{ \g }{^0^0} \tnabla _0 \tnabla _0 \A_0
 +
  \t{ \g }{^i^j} \tnabla _0 \tnabla _i \A_j
    \underbrace{=}_{\mbox{\scriptsize by \eqref{5X22.3}, \eqref{8V22.31} and \eqref{8V22.32b}}} 0
\,.
\label{2V22.3a}
\end{eqnarray}

Commuting $\g^{\alpha\nu}\tnabla _\alpha$ with $\tBox_\g$ and using \eqref{5X22.3} gives
\begin{equation}
	\Box_\g ( \g^{\alpha\beta} \tnabla_\alpha \A_\beta  )  =
 \redmusquared \g^{\alpha\beta} \tnabla_\alpha \A_\beta + \g^{\rho \sigma}\tnabla_\rho(( \g^{\alpha\beta}\tnabla_\alpha \A_\beta)\tfrac{1}{n} \tnabla_\sigma n)
\,.
\label{2V22.6c}
\end{equation}
So $\tnabla^\alpha \A_\alpha$ satisfies the wave equation \eqref{2V22.6c} with vanishing Cauchy data,
and uniqueness of solutions gives \eqref{eq:constraint}.

Summarising, we have proved:

\begin{proposition}
 \label{P8V22.1}
A field $A_\alpha$ satisfies \eqref{eq:fieldeq} with $\m ^{\alpha\beta} =\g^{\alpha\beta}$ if and only if the Cauchy data are constrained by equations \eqref{8V22.31} and \eqref{8V22.32b} on $\hyp$, and then it can be obtained by solving the wave equation \eqref{5X22.3}.
\end{proposition}

\section{The case $\m = g + \redp u\otimes u$}

The case $\m\ne \g$ turns-out to be much more involved computationally, and it is not clear whether a consistent Cauchy problem can be obtained for general spacetimes. The arising complications hint strongly to the fact that the case $\m=\g$ provides a better model for the problem at hand. As a partial contribution to the well-posedness question we analyse the case where $g$ is flat, the medium is moving inertially, $\partial_\mu u^\alpha=0$, and the refractive index $n$ is constant.

As already mentioned in the introduction, since there is no unique extension of the Proca mass-term from vacuum to dielectric media  we consider  the more general case
\begin{equation}
	\m^{ \alpha \beta } = g^{ \alpha \beta } + \redp \, u^\alpha u^\beta
	\,,
  \label{15IV23.1}
\end{equation}
with $\redp \in \R$. It turns out that the equation satisfied by $A_0$ is hyperbolic only for  $\redp<1$ (cf.\ \eqref{27XI22.1}-\eqref{27XI22.1b} below). For this reason we restrict our analysis to $\redp < 1$.

The overall strategy is the same as in the previous section: derive a system of wave equations which will be used to evolve $A_\mu$, see \eqref{27XI22.1}-\eqref{27XI22.4} below, and show that suitable constraints on the Cauchy data  lead to a solution of the problem at hand.

We start by noting that, in the current setting, Equation \eqref{5X22.2} simplifies to
\begin{equation}
	\Box_\g A_\beta - \t{ \g }{^\sigma^\rho} \nabla _\beta \nabla _\sigma A_\rho - \redmusquared A_\sigma \t{ \m }{^\sigma^\rho} \t{ \g }{_\rho_\beta} = 0
\,,
\label{eq:wave}
\end{equation}
with $\Box_\g \equiv \g^{\alpha \beta} \nabla_\alpha \nabla_\beta$.
Equation \eqref{eq:constraint} reads now
\begin{equation}
	\nabla _\alpha \left(
 \t{ \m }{^\alpha^\beta} A_\beta \right) = 0
\,,
 \label{24IV23.21}
\end{equation}
while \eqref{10III22.3cd} becomes
\begin{equation}
  \redmusquared \t{ \m }{^\alpha^0} A_\alpha
    =
    \nabla _i \t{ G }{^i^0}
     \,.
\label{eq:CauchyConstraint}
\end{equation}
Yet another equation can be obtained by contracting   $\m^{\beta \alpha} \nabla_\alpha$ with   \eqref{eq:wave} and using \eqref{24IV23.21}:
\begin{equation}
	\Box_\m \nabla_\alpha \left( \t{ \g }{^\alpha^\beta} A_\beta \right) +  \redmusquared \nabla _\alpha \left( \t{ \m }{^\alpha^\sigma}\t{ \g }{_\sigma_\rho}\t{ \m }{^\rho^\beta} A_\beta \right)= 0
\,.
\label{2V22.6a}
\end{equation}
In the case $\m=\g$ (equivalently,  $\redp=1-n^2$)  this condition was  trivially fulfilled by solutions of the field equations, but it isn't anymore if $\redp  \ne 1-n^2$. In what follows the last condition will be assumed.

Using \eqref{15IV23.1} and the definition of the Gordon metric \eqref{eq:Gordon}, Equation~\eqref{2V22.6a} can be rewritten as
\begin{equation}
	\Box_\m \p_0 A_0  - \redmusquared n^{-2}(1 - \redp)\p_0 A_0 = 0
\,.
\label{28XI22.1}
\end{equation}
 Defining
\begin{equation}
	\varphi := \p_0 A_0
	\,,
	\label{28XI22.2}
\end{equation}
we are thus led  to analyse the following system of equations:
\begin{eqnarray}
   (\Box_\m   - \redmusquared  n^{-2} (1 - \redp) ) A_0 &=&  0
    \,,
    \label{27XI22.1}
    \\
    (\Box_\g - \redmusquared) A_i &=&  (1-\redp - n^2) \p_i \varphi
    \,,
    \label{27XI22.3}
    \\
     \Box_\m \varphi &=& \redmusquared n^{-2} (1 - \redp) \varphi
    \,.
     \label{27XI22.4}
\end{eqnarray}

Note that the operator $\Box_\m$ is  hyperbolic when $\redp\ne 1$,  $3d$  elliptic when $\redp=1$, and $4d$-elliptic when $\redp>1$:
\begin{equation}
	\Box_\m = (\redp-1) \partial_0^2 + \partial_1^2 + \partial_2^2 + \partial_3^2
\,.
\label{27XI22.1b}
\end{equation}
For this reason we will only consider $\redp <1$; and recall that $\redp = 1-n^2$  has already been covered by Proposition~\ref{P8V22.1}.

\begin{proposition}
 \label{p25IV23.1}
Solutions of \eqref{eq:wave} with $\redp < 1$, $\redp \ne 1-n^2$, are uniquely parameterised by two arbitrary fields $\vec A|_{t=0}$ and $\dot {\vec A}|_{t=0}$.
In fact,
a field $A_\alpha$ satisfies \eqref{eq:wave} if and only if \eqref{24IV23.21}-\eqref{eq:CauchyConstraint}  hold at $t=0$, and it can be obtained by solving the system of wave equations \eqref{27XI22.1}-\eqref{27XI22.4}.
\end{proposition}

\proof

The  constraint \eqref{eq:CauchyConstraint} shows that for each $t$ the field $A_0(t,\cdot)$ must solve the elliptic equation
\begin{equation}\label{24XI22.1}
  ( \Delta   - (1- \redp) n^{-2} \redmusquared ) A_0|_{t=0} = \partial_i \dot A^i |_{t=0}
   \,,
\end{equation}
where a dot denotes a time derivative.

We thus let  $ A_0|_{t=0}$ be the solution of \eqref{24XI22.1}
 (which can be found with the help of the usual Green function for the Helmholtz equation),
and set
\begin{equation}\label{24XI22.2}
       \partial_0 A_0|_{t=0}: = \frac{1}{1-\redp} \p_i A_i |_{t=0},
\end{equation}
as necessary for \eqref{eq:constraint}.

The initial data for $\varphi$ is set to be
\begin{align}
	\varphi|_{t=0} &= \frac{1}{1 - \redp} \p_i A_i |_{t=0}
	\,,
	\label{28XI22.5}
	\\
	 \dot \varphi|_{t=0} &= \frac{1}{1 - \redp} \p_i \dot A_i |_{t=0}
	 \,.
	\label{28XI22.6}
\end{align}

It remains to show that the set of evolution equations \eqref{27XI22.1}-\eqref{27XI22.4} propagates the constraint \eqref{24XI22.2} from the initial data.

First, the condition $\varphi |_{t=0} = \p_0 A_0 |_{t=0}$ clearly holds. Further,
\begin{align}
	\p_0 ( \varphi - \p_0 A_0 )|_{t=0} &= \frac{1}{1 - \redp} (\p_i \dot A_i) |_{t=0} - \p_0^2 A_0 |_{t=0}
	\notag
	\\
	&= \frac{1}{1 - \redp}  \left[( \Delta   - (1- \redp) n^{-2} \redmusquared ) A_0 - (1 - \redp ) \p_0^2 A_0\right] |_{t=0}
	\notag
	\\
	&= \frac{1}{1 - \redp}  ( \Box_\m  - (1- \redp) n^{-2} \redmusquared ) A_0|_{t=0}
	\notag
	\\
	&=0
	\,,
\end{align}
and
\begin{align}
	\Box_\m ( \varphi - \p_0 A_0 ) &=  \redmusquared n^{-2} (1 - \redp) ( \varphi - \p_0 A_0)\,,
\end{align}
and thus via the vanishing initial data
\begin{equation}
	\varphi \equiv \p_0 A_0
	\,.
	\label{27IV23.1}
\end{equation}

We continue by noting that
\begin{equation}
	\Box_\g = \Box_\m + (1 - \redp - n^2) \p_0^2
	\,.
\label{28XI22.7}
\end{equation}
This, together with \eqref{27IV23.1} and the time-derivative of \eqref{27XI22.1}, implies that
\begin{equation}
	\Box_\g \p_0 A_0 = n^{-2} (1- \redp) \redmusquared \p_0 A_0 + (1-\redp - n^2) \p_0^2 \varphi
	\,.
\label{28XI22.8}
\end{equation}
Taking the divergence of \eqref{27XI22.3} leads to
\begin{equation}
   \Box_\g \p_i A^i    =  \redmusquared \p_i A^i  + (1-\redp - n^2)\Delta  \varphi
   \,.
    \label{28XI22.9}
\end{equation}
Subtracting \eqref{28XI22.8} from \eqref{28XI22.9} and using \eqref{27XI22.4} as well as \eqref{27IV23.1}  one obtains
\begin{equation}\label{28XI22.9.2}
  \Box_\g(-(1 - \redp )\partial_0  A_0 + \partial_i A^i) = \redmusquared  (-(1 - \redp )\partial_0  A_0 + \partial_i A^i)
   \,.
\end{equation}
As before, the Cauchy data for this wave equation vanishes. Indeed, \eqref{24XI22.2} explicitly states that
\begin{equation}\label{28XI22.10}
	 ((1 - \redp) \p_0 A_0 - \p_i A^i )|_{t=0} = 0
	\,.
\end{equation}
Next, using \eqref{27XI22.1} and \eqref{24XI22.1} yields
\begin{align}\label{28XI22.11}
	\p_0 ((1 - \redp )\p_0 A_0 - \p_i A^i)|_{t=0}
		&=0
		\,.
\end{align}
Uniqueness of solutions of the Cauchy problem for \eqref{28XI22.9.2} together with \eqref{28XI22.10}-\eqref{28XI22.11} shows that we have
\begin{equation}\label{28XI22.12}
 \partial_0  A_0    \equiv \varphi \equiv \frac{1}{1 - \redp} \p_i A_i
\end{equation}
everywhere.

Since $\varphi = (1-\redp - n^2)^{-1} \g^{\sigma \rho} \p_\sigma A_\rho$ due to \eqref{28XI22.12}, the equivalence of \eqref{27XI22.1}-\eqref{27XI22.4} to \eqref{eq:wave} is immediate.

\qed

\section*{Acknowledgements}
Many useful discussions with Thomas Mieling are acknowledged.
Research supported in part by the Austrian Science Fund (FWF), Project P34274 and by the European Union (ERC, GRAVITES, project no 101071779). Views and opinions expressed are however those of the authors only and do not necessarily reflect those of the European Union or the European Research Council Executive Agency. Neither the European Union nor the granting authority can be held responsible for them.

\providecommand{\bysame}{\leavevmode\hbox to3em{\hrulefill}\thinspace}
\providecommand{\MR}{\relax\ifhmode\unskip\space\fi MR }
\providecommand{\MRhref}[2]{%
  \href{http://www.ams.org/mathscinet-getitem?mr=#1}{#2}
}
\providecommand{\href}[2]{#2}

\end{document}